\documentclass[aps,prl,twocolumn,floats]{revtex4}

\usepackage{graphicx}

\begin{document}

\title{Distance-$d$ covering problems in scale-free networks with degree 
correlations}

\author{$^{1}$ Pablo Echenique, $^{1}$ Jes\'us G\'omez-Garde\~nes,$^{1}$
Yamir Moreno, $^{2}$ Alexei V\'azquez}

\affiliation{$^{1}$Institute of Biocomputation and Physics of Complex
Systems (BIFI), University of Zaragoza, Zaragoza 50009, Spain}

\affiliation{$^{2}$Department of Physics, University of Notre Dame, IN
46556, USA}

\date{\today}

\begin{abstract}

A number of problems in communication systems demand the distributed
allocation of network resources in order to provide better services,
sampling and distribution methods. The solution to these issues is
becoming more challenging due to the increasing size and complexity of
communication networks. We report here on a heuristic method to find
near-optimal solutions to the covering problem in real communication
networks, demonstrating that whether a centralized or a distributed design
is to be used relies upon the degree correlations between connected
vertices. We also show that the general belief that by targeting the hubs
one can efficiently solve most problems on networks with a power law
degree distribution is not valid for assortative networks.

\end{abstract}

\maketitle 

\bibliographystyle{apsrev}

The allocation of network resources to satisfy a given service with the
least use of resources, is a frequent problem in communication networks.
For instance, a highly topical problem is the developing and deploying of
a digital immune system to prevent technological networks from virus
spreading. In this case, it is worthwhile to characterize whether a
centralized organization or a distributed approach is the best choice
\cite{white}. Clearly, this is the first decision, and perhaps one of the
fundamental ones, that must be taken before proceeding with other
technical issues.  Another natural ground includes the placement of web
mirror servers. The solution to such problems is becoming more challenging
due to the increasing size of social and technological networks. Heuristic
approaches that provide hints and pave the way for more elaborated
strategies would be welcome.  For this purpose we must identify which
individuals are the ideal candidates to transmit, collect, monitor or
prevent information and virus spreading across the net
\cite{white,ba00,pv01c,db01a,alph01,cohen03}.

The solution to this and similar problems may be computationally easy or
hard depending on the topological properties of the underlying graph
\cite{gj79,martin01,mezard02,wh00b,vw02}. In particular,
communication and many other real-world graphs are characterized by wide
fluctuations in the vertex degrees \cite{ab01a,dm002b,newman03a}, where
the degree of a vertex is the number of edges attached to it. This means
that, in addition to a high number of small degree vertices, there are
hubs connected to a large number of other vertices. The existence of hubs
has been exploited to develop strategies aimed at enhancing network
resilience to damage \cite{ba00}, virus spreading
\cite{pv01c,db01a,cohen03}, and searching algorithms \cite{alph01}.
Additionally, real-world networks are characterized by degree correlations
between connected vertices \cite{pvv01,n02a}. These degree correlations
have been shown to affect the computational complexity of hard problems on
graphs with wide fluctuations in the vertex degrees \cite{vw02}.

We report here on a heuristic method that allows us to find near-optimal
solutions to the covering problem in real-world networks. Specifically, we
are interested in the problem of computing the minimum set of covered
vertices (referred to henceforth as servers) such that every vertex is
covered or has at least one covered vertex at a distance at most $d$
(distance-$d$ covering problem), where the distance between two vertices
in the graph is the minimum number of hops necessary to go from one vertex
to the other. Each server will then provide service to or monitor those
vertices within a distance $d$.  Using a heuristic algorithm that targets
high degree vertices, we compute an upper bound to the minimum fraction of
servers needed to cover these graphs. We find out that the solution to the
distance-$d$ covering problem strongly depends on the degree of similarity
between the connected vertices. As a consequence, we show that when
designing networked systems, whether a centralized or distributed design
is to be used relies upon the network properties at a local level. Our
primary intent is not to develop an optimal algorithm. Instead, our main
focus is in assessing the impact of correlations on the design of
networked systems, and hence provide motivations, or lack thereof, for
moving to more complex heuristics in the context of covering problems in
real nets.

\begin{figure} 
\begin{center} 
\includegraphics[width=3in]{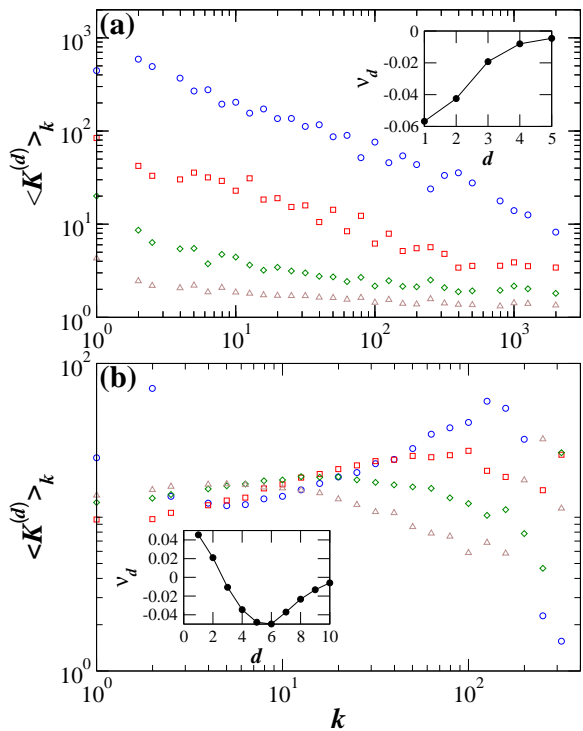}
\end{center} 

\caption{Average degree $<K^{(d)}>_k$ of the distance-$d$
neighbors of a vertex with degree $k$, for $d=1$ (circles), $d=2$
(squares), $d=3$ (diamonds) and $d=4$ (triangles). Note that the
average neighbor degree introduced in Ref. \cite{pvv01} corresponds
with $<K^{(1)}>_k$. (a) $<K^{(d)}>_k$ vs $k$ for the AS
graph. The inset shows the exponent $\nu_d$ obtained from the best fit
to the power law $<K^{(d)}>_k=Ak^{\nu_d}$ in the range
$k>1$. Similar results are obtained for the Gnutella graph, but with
more fluctuations due to its small size. (b)$<K^{(d)}>_k$ vs $k$
for the Router graph. The inset shows the exponent $\nu_d$ obtained
from the best fit to the power law $<K^{(d)}>_k=Ak^{\nu_d}$ in
the range $10\leq k\leq100$.}

\label{fig1} 
\end{figure}

The communication networks considered in this work are the
following. AS: Autonomous system level graph representations of the
Internet as of April 16th, 2001.  Gnutella: Snapshot of the Gnutella
peer to peer network, provided by Clip2 Distributed Search
Solutions. Router: Router level graph representations of the
Internet. All these graphs are sparse with an average degree around 3,
small worlds \cite{ws98} with an average distance between vertices
less than 10, and they are characterized by a power law degree
distribution $p_k\sim k^{-\gamma}$, with $\gamma\approx2.2$. A
detailed characterization of these graphs is presented in Refs.
\cite{vazquez03} (Gnutella) and \cite{pvv01,vpv02b,pv04} (AS and
Router graphs). They differ, however, in their degree correlations
between nearest neighbor vertices. The AS and Gnutella graphs exhibit
disassortative degree correlations, with a tendency to have
connections between vertices with dissimilar degrees
(Fig. \ref{fig1}a).  In contrast, the Router graph displays
assortative degree correlations, with a tendency to establish
connections between vertices with similar degrees (Fig.
\ref{fig1}b). In this work we are interested in covering problems
beyond $d=1$, therefore we also analyze the degree correlations for
$d>1$ \cite{note}. For the disassortative graphs, the average degree of
distance-$d$ neighbors $<K^{(d)}>_k$, restricted to root vertices
with degree $k$, follows the same trend as $<K^{(1)}>_k$, tending
to be less correlated for larger $d$ (Fig. \ref{fig1}a).  For the
assortative graph, however, the degree correlations are assortative up
to $d=2$, becoming disassortative for $d>2$
(Fig. \ref{fig1}b). Finally, for $d>6$ the degree correlations in the
originally assortative graph show a similar trend than in the
disassortative graphs.

We propose the following heuristic algorithm to obtain an upper bound
to the distance-$d$ covering problem. {\it Local algorithm}: For every
vertex in the graph, cover the highest degree vertex at a distance at
most $d$ from the vertex. In case there is more than one vertex with
the highest degree, one of them is selected at random and covered. To
test this algorithm we first consider the case $d=1$, known as the
dominating set problem \cite{gj79}. In this case we can use a
leaf-removal algorithm as a reference method, which yields a nearly
optimal solution together with an error estimate \cite{lvw}. The
leaf-removal algorithm is defined as follows. To each vertex $i$ we
assign two state variables $x_i$ and $y_i$, where $x_i=0$ ($x_i=1$) if
the vertex is uncovered (covered) and $y_i=0$ ($y_i=1$) if the vertex
is undominated (dominated). Here a vertex is said to be dominated if
it has at least one neighbor covered. Starting with all vertices
uncovered and undominated ($x_i=y_i=0$ for all $i$), iteratively,
({\it i}) select a vertex with degree one (leaf). If it is not
dominated, cover its neighbor, set dominated its second neighbors, and
then remove the leaf, its neighbor, and all their incident
edges. ({\it ii}) If no vertex with degree one is found, then cover
the vertex with the larger degree (hub), set dominated its neighbors,
and then remove the hub and all its incident edges. Finally, if some
vertices with degree zero remain, they are covered if they are not
dominated, and removed from the graph.  Since step ({\it i}) always
provides an optimal solution, the error in computing the average
fraction of covered vertices $\left<x\right>=\sum_{i=1}^Nx_i/N$ is
less than or equal to the fraction of vertices covered applying step
({\it ii}).

\begin{figure}
\begin{center}
\includegraphics[width=3.0in]{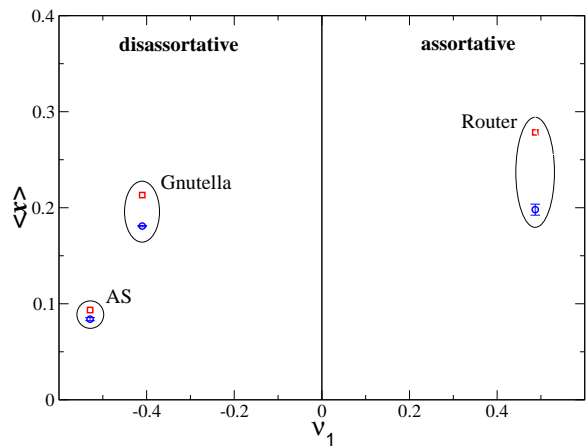}
\end{center}

\caption{Average fraction of servers $\left<x\right>$ needed to cover a
graph under the constraint that a vertex should have a server at most at a
distance $d=1$, using the leaf-removal (circles) and local (squares)
algorithms, as a function of the exponent $\nu_1$ defined in Fig.  
\ref{fig1}, with negative and positive values corresponding to
disassortative and assortative graphs, respectively.}

\label{fig2} 
\end{figure}

The comparison between the local and leaf-removal algorithms is shown
in Fig. \ref{fig2}. First, notice that the solutions obtained with the
leaf-removal algorithm are almost exact for the networks considered
here and $d=1$.  The local algorithm yields satisfactory, though
non-optimal, solutions to the covering problem, with some differences
depending on correlations between connected vertices. For the AS and
the Gnutella graphs, which exhibit disassortative degree correlations,
the local algorithm gives a good estimate, quite close to the optimal
one for the AS graph.  In contrast, for the Router graph we observe a
larger deviation from the optimal solution. The origin of this
difference is due to the fact that the local algorithm exploits the
degree fluctuations among connected vertices. Indeed, these
fluctuations are bigger in disassortative graphs as connected vertices
likely have different degrees.  In contrast, in assortative graphs,
although there may be high degree fluctuations between two vertices
selected at random, connected vertices tend to have similar degrees,
resulting in poorer solutions. These results indicate that the general
belief that heuristic algorithms targeting the hubs may be sufficient
to solve computational problems on graphs with wide degree
fluctuations may not be the case for assortative graphs.

\begin{figure}
\begin{center}
\includegraphics[width=3.0in]{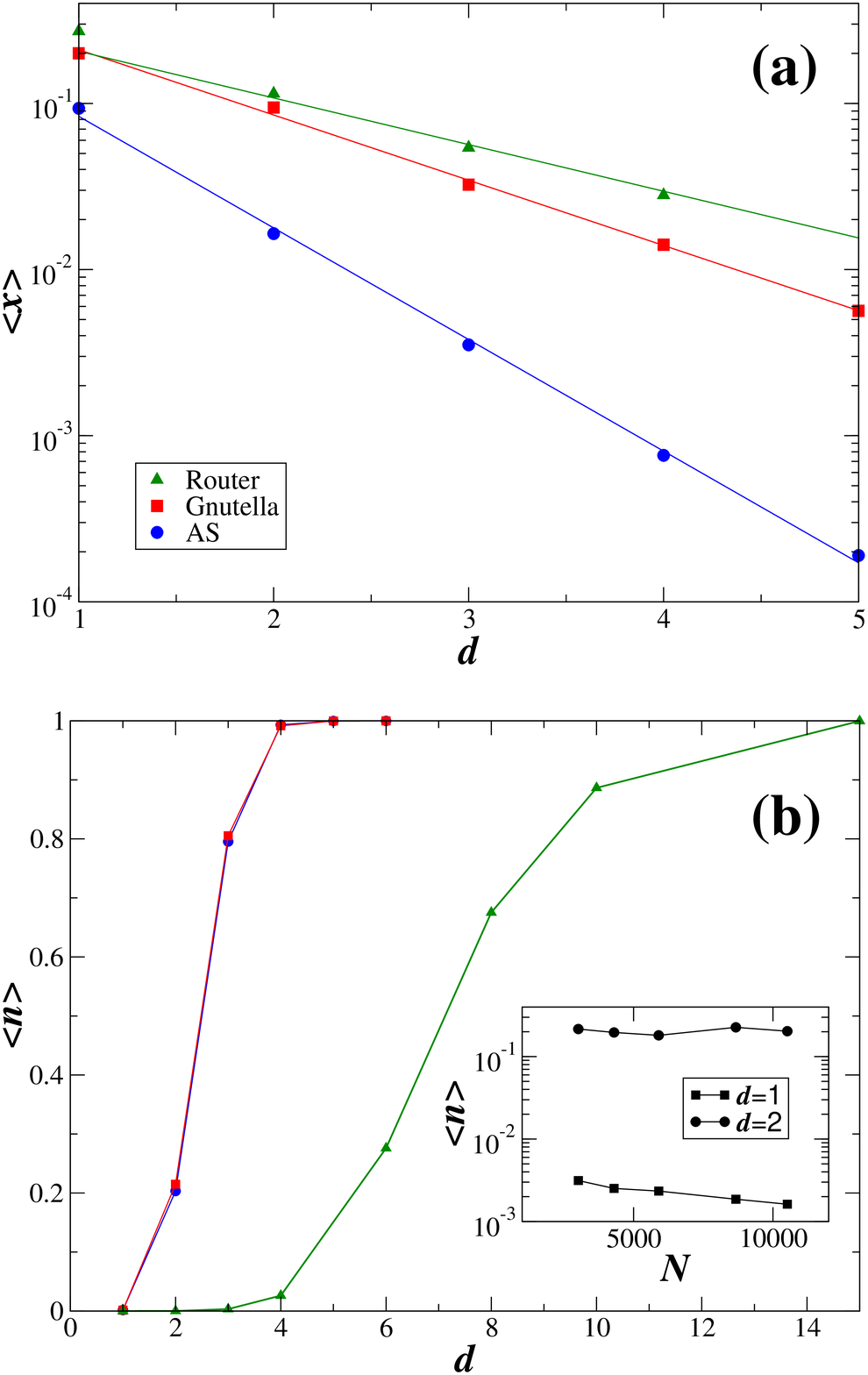}
\end{center}

\caption{(a) Average fraction of servers $\left<x\right>$ covering the
graph for different values of $d$.The continuous lines are the best fits
to an exponential decay. (b) Average fraction of vertices $\left<n\right>$
served by a server for different values of $d$. The inset shows the graph
size dependence of $\left<n\right>$ for the AS graph and $d=1,2$.}

\label{fig3}
\end{figure}

The $d=1$ covering problem results in a distributed architecture because a
finite fraction of the vertices is covered. Let us now extend the method
and discuss the results obtained with the local algorithm for the more
general and complex problem $d>1$. In Fig. \ref{fig3}a we show that, with
increasing $d$, the average fraction of servers decays exponentially fast,
indicating that if we allow the servers to be more distant, a substantial
decrease in the number of required servers is obtained. This exponential
decay is a consequence of the small-world property of these networks,
characterized by an average distance between vertices that grows as or
slower than the logarithm of the number of vertices.  The decrease in
$\left<x\right>$ is, however, achieved at the expense of an increase in
the average fraction of vertices $\left<n\right>$ covered by a server
(Fig. \ref{fig3}b). This is a key metric as it marks the trade-off
between the number of servers needed and their capacity.

Again, a remarkable difference depending on the graph assortativities is
appreciated. For the Gnutella and AS graphs, with disassortative
correlations, $\left<n\right>$ increases significantly from $d=1$ to
$d=2$. Indeed, a finite size study for the AS graph, with a growing
tendency from 1997 to 2002 \cite{pvv01}, reveals that $\left<n\right>$
decreases to zero with increasing the graph size for $d=1$, while it
remains almost constant for $d=2$ or larger (see inset of Fig.
\ref{fig3}b). On the other hand, in the Router graph, with assortative
correlations, $\left<n\right>$ increases much slower with increasing $d$,
being almost zero up to $d=3$ (Fig. \ref{fig3}b). These results are the
signature of a phase transition. There is a threshold distance $d_c$ such
that the average fraction of vertices served by a covered vertex is very
small for $d\leq d_c$, going to zero with increasing $N$, while it is
finite for $d>d_c$. For disassortative graphs $d_c=1$ while for
assortative ones $d_c>1$. Note that the value $d_c\approx3$ for the Router
graph coincides with the distance where the degree correlations become
disassortative, indicating that the phase transition is determined by the
change in the degree correlations. Furthermore, this transition gives a
practical measure to get the desired trade-off between $\left<x\right>$
and $\left<n\right>$.

\begin{figure}
\begin{center}
\includegraphics[width=3in]{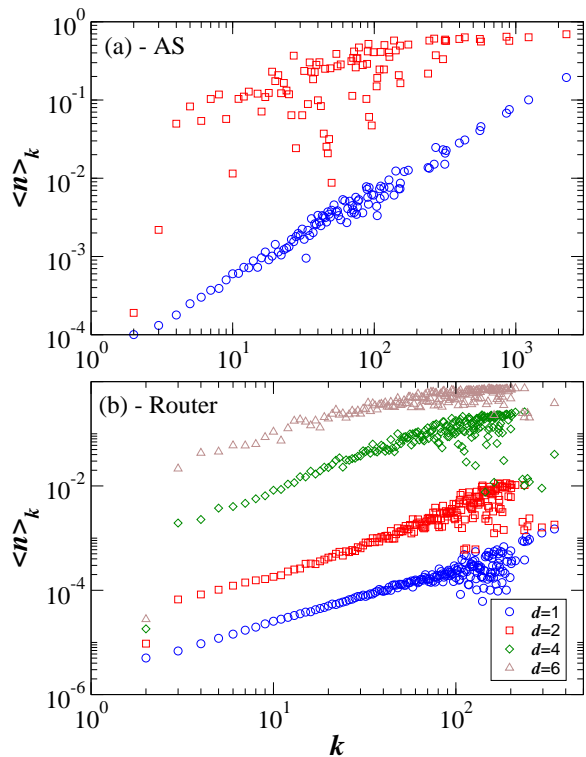}
\end{center}

\caption{Average number of covered vertices (servers) $\left<n\right>_k$
restricted to vertices with the same degree $k$ for several values of $d$.
The figures show that for disassortative graphs ({\bf a}), the servers
should have a large capacity to serve a finite fraction of the graph even
for small to moderate values of $d$. On the contrary, for assortative
graphs ({\bf b}), the fraction of servers is a negligible fraction of $N$
up to large values of $d$.}

\label{fig4}
\end{figure}

Since the graphs considered here are characterized by wide fluctuations in
the vertex degrees, we have also computed the average number of covered
vertices $\left<n\right>_k$, restricted to vertices with the same degree
$k$. In all cases we observe an increasing tendency of $\left<n\right>_k$
with $k$, as it is expected from the definition of the local algorithm,
which targets high degree vertices. Two distinct behaviors are once again
observed depending on the degree correlations. In the disassortative
graphs, $\left<n\right>_k$ is already as large as 10\% of the vertices for
$d=2$ and $k>10$ (Fig. \ref{fig4}a). In contrast, in the assortative
graphs, only beyond $d=4$, one observes that large value of
$\left<n\right>_k$.

The striking differences between disassortative and assortative
correlations have important consequences regarding how resources are
allocated. For disassortative graphs, except for the case $d=1$, one
would need servers with a vast capacity, covering a large fraction of
vertices. The most efficient strategy is, therefore, the allocation of
resources in a few servers with a large capacity. The scalability of
the server system would in this case be determined by the single
server capacities, which should be increased as the graph size
grows. In the assortative case, we have a different scenario. The
decrease of the number of servers with increasing $d$ is not as
dramatic as for the disassortative graphs. In compensation, each
server covers a small fraction of vertices. Hence, the most efficient
strategy is to allocate the resources in a large number of servers
with a limited capacity. The scalability of the system would be driven
by the number of required servers, which augments with increasing the
graph size.  In turn, regarding the design of communication networks,
we can decide between disassortative or assortative topologies
depending on the available resources. A disassortative topology will
be more appropriate for a centralized design, with a few servers
having a large capacity, while an assortative network will be best
suited for a distributed design, when a large number of servers have a
limited capacity.

It is worth stressing that the heuristic proposed is based on a local
knowledge of the network (only requiring information about the graph
topology up to a distance $d$), a key property of utmost importance
for most real applications. Indeed, all the graphs considered here are
incomplete representations of the systems they are aim to represent
\cite{traceroute04}, as it generally happens in graph representations
of large systems.

Finally, the present study shows that the general belief that by targeting
the hubs one can efficiently solve most problems on networks with a power
law degree distribution (percolation, spreading, searching, covering, etc)
is not valid if the degree correlations are assortative. This
conclusion is of special relevance in the analysis of social systems where
assortative networks are the general rule. Furthermore, we have shown that
whether the degree correlations are assortative or disassortative may
depend on the distance between the connected vertices, indicating that
different strategies may be used depending on the characteristic distance
of the covering problem.

\begin{acknowledgments}
P.\ E.\ and J.\ G-G\ acknowledge financial support of the MECyD
through FPU grants. Y.\ M.\ is supported by a BIFI Research
Grant. This work has been partially supported by the Spanish DGICYT
projects BFM2002-01798 and BFM2002-00113.
\end{acknowledgments}


\begin{thebibliography}{23}
\bibitem{note} Note that while other topological properties may affect the
performance of the algorithm they would do it through changes in the
degree correlations.
\end{thebibliography}
\end{document}